\documentclass[letterpaper]{article}
\pdfoutput=1
\usepackage{jheppub}
\usepackage{amsmath}
\usepackage{graphicx}
\usepackage[config=altsf]{subfig}
\usepackage{feynmf}
\usepackage{siunitx}
\usepackage{array}
\usepackage{slashed}
\usepackage{afterpage}
\usepackage{appendix}
\usepackage{multirow}

\title{Small Vacuum Energy from Small Equivalence Violation\\
  {\begin{center} in Scalar Gravity\end{center}}}
  \author[a]{Prateek Agrawal}
\author[b]{and Raman Sundrum}
\affiliation[a]{
  Department of Physics, Harvard University,
 Cambridge, MA 02138, USA}
\affiliation[b]{
  Department of Physics, University of Maryland,
 College Park, MD 20742, USA}
\emailAdd{prateekagrawal@fas.harvard.edu}
\emailAdd{raman@umd.edu}

\abstract { The theory of scalar gravity proposed by Nordstr\"{o}m,
and refined by Einstein and Fokker, provides a striking analogy to
general relativity.  In its modern form, scalar gravity appears as the
low-energy effective field theory of the spontaneous breaking of
conformal symmetry within a CFT, and is AdS/CFT dual to the original
Randall-Sundrum I model, but without a UV brane.  Scalar gravity
faithfully exhibits several qualitative features of the cosmological
constant problem of standard gravity coupled to quantum matter, and
the Weinberg no-go theorem can be extended to this case as well.
Remarkably, a solution to the scalar gravity cosmological constant
problem has been proposed, where the key is a very small violation of
the scalar equivalence principle, which can be elegantly formulated as
a particular type of deformation of the CFT.  In the dual AdS picture
this involves implementing Goldberger-Wise radion stabilization where
the Goldberger-Wise field is a pseudo-Nambu Goldstone boson.  In
quantum gravity however, global symmetries protecting pNGBs are not
expected to be fundamental.  We provide a natural six-dimensional
gauge theory origin for this global symmetry and show that the
violation of the equivalence principle and the size of the vacuum
energy seen by scalar gravity can naturally be exponentially small.
Our solution may be of interest for study of non-supersymmetric CFTs
in the spontaneously broken phase.

}

\preprint{UMD-PP-017-012}

\begin{document}
\maketitle 
\section{Introduction} 

The cosmological constant problem
(see~\cite{Weinberg:1988cp,Polchinski:2006gy}
for reviews) is notoriously challenging from the viewpoint of general
relativity effective field theory as well as quantum gravity.
Therefore, any simplified theory which replicates significant
qualitative features of the problem is very valuable. In this paper,
we will focus on such an analog~\cite{Sundrum:2003yt}, the theory of
scalar gravity. This theory, first introduced by
Nordstr\"{o}m~\cite{1913AnP...347..533N}, was perfected and formulated
as a theory of curved spacetime by Einstein and
Fokker~\cite{Einstein:1914bu}, and possesses a very faithful version
of the equivalence principle.
In the modern era, this theory emerged  as the low-energy effective
field theory
of the spontaneous breaking of conformal symmetry within a conformal
field theory (CFT)~\cite{Isham:1970gz,Isham:1971dv}.
The associated Goldstone boson, the ``dilaton'', mediates a scalar
``gravitational''
force and couples to the trace of the stress energy tensor of any
other light degrees of freedom.  
The AdS/CFT dual of this appears in the original Randall-Sundrum I
model~\cite{Randall:1999ee}, where the dilaton is dual to the radion,
and light
matter is present on the IR brane.  More precisely, to have only
scalar gravity
present, the UV brane is removed.

Reference~\cite{Sundrum:2003yt} studied the coupling of quantum matter
to scalar gravity and showed that there was a very faithful version of
the cosmological constant problem (CCP) in this theory.  Within the
grammar
of scalar gravity the problem seems as robust as in general relativity
(tensor gravity) and indeed the Weinberg no-go
theorem~\cite{Weinberg:1988cp} can be extended
to this case as well.
It is therefore of considerable interest that a solution to the scalar
version of the CCP has been proposed \cite{CPR,Coradeschi:2013gda}. 
The key to this proposal
is a very small violation of the scalar
equivalence principle, which is technically considerably easier to
achieve and to deeply understand than such a violation in general
relativity. Indeed, in the AdS/CFT dual picture, it just involves a
particular implementation of the Goldberger-Wise (GW) radion
stabilization mechanism~\cite{Goldberger:1999uk} where the GW field is a
pseudo-Nambu Goldstone Boson (pNGB). In quantum gravity, global
symmetries such as those associated to
pNGBs are not expected to be fundamental, and should be realized as
accidental symmetries enforced by gauge structure. In this paper, we
realize the mechanism of reference~\cite{CPR} in just such a manner,
and show that the violation of the equivalence principle and the
size of the cosmological constant can naturally be
exponentially small. We will formulate our refined mechanism in the
AdS-dual language, as a six-dimensional warped effective field theory.

To be self-contained, we will review scalar gravity, its CCP and the
proposed solution, closely following the discussion of
refs.~\cite{Sundrum:2003yt,Coradeschi:2013gda}.

\subsection{Scalar gravity}

Even though scalar gravity is ruled out experimentally (for example,
light does not bend in a scalar gravity field since the stress-energy
tensor of a free Maxwell field is traceless), it serves as a useful
analogy to general relativity. A key feature is that the scalar
gravity theory respects the Strong Equivalence Principle.  This
implies that the inertial and gravitational mass of compact objects is
the same, with the strong version including scenarios where the
gravitational binding energy of the object is not
negligible~\cite{Deruelle:2011wu}.

The covariance of scalar gravity is explicitly seen in the metric
formulation,
\begin{align}
  g_{\mu\nu}
  &=
  \frac{\varphi^2}{M_{pl}^2}
  \eta_{\mu\nu} \,,
  \label{eq:metric}
\end{align}
where $\varphi$ is the scalar graviton, and $M_{pl} = \langle \varphi
\rangle$ is the scalar gravity Planck scale.
This equation does not seem covariant, but can be viewed as a
diffeomorphism gauge-fixed version of the 
generally covariant constraint of the vanishing of the Weyl
tensor\footnote[2]{In quantum effective field theory,
this constraint can be imposed at the level of the path integral using
a Lagrange
multiplier.
},
\begin{align}
  C_{abcd}
  &=
  R_{abcd}
  -\left( g_{a[c} R_{d]b} - g_{b[c}R_{d]a} \right)
  +\frac13
  R g_{a[c} g_{d]b}
  =0
  \,.
  \label{eq:weyl}
\end{align}
In
$d\geq4$, it is a necessary and sufficient condition for obtaining a
conformally flat solution that the Weyl tensor vanish. (Here we
restrict to $d=4$.) Then, a
modified 
Einstein-Hilbert action serves as the action for scalar gravity
\begin{align}
  S
  &=
  \int d^4 x \sqrt{g} \left(
  \frac{M_{pl}^2}{12} R + \Lambda + \mathcal{L}_{matter}
  \right)
  \,.
  \label{eq:scalar-gravity}
\end{align}
Note, the sign of the Ricci scalar is opposite of that in the usual
Einstein-Hilbert action (which is negative in the
$(+---)$ signature we use). 
This is a consequence of the fact that
in GR the conformal mode has the ``wrong-sign kinetic term'', which is
not a
problem since this mode is non-propagating. In the case of scalar
gravity, however, this
is
precisely the only propagating mode allowed by the Weyl constraint and
hence the sign of the Einstein-Hilbert action must be reversed.
 In 
particular, we can recover scalar gravity equation of motion
in the Jordan frame,
\begin{align}
  R = \frac{12}{M_{pl}^2}T^\mu_\mu
  \,.
  \label{eq:nordstrom}
\end{align}
Notice that a cosmological constant term $\Lambda$ naturally appears 
in equation \eqref{eq:scalar-gravity} and will be renormalized by
quantum matter.

In its modern incarnation, this scalar gravity theory can be viewed as
the low-energy limit of a
CFT with spontaneously broken conformal invariance.
The subgroup of diffeomorphisms which preserve the form of the metric
as in equation~\eqref{eq:metric} are conformal transformations,
\begin{align}
  x &\to x'(x) \,,\\
  g_{\mu\nu}(x) &\to f^2(x) g_{\mu\nu}(x)\,,
\end{align}
or
\begin{align}
  \varphi (x) \to f(x) \varphi(x) \,.
\end{align}
Rewriting $\varphi(x)$ as $e^{\tau(x)}$, for a constant scaling
$e^\lambda$ we get a shift, 
\begin{align}
  \tau(x) \to \tau(\lambda x) + \lambda
  \,.
  \label{eq:dilaton-transform}
\end{align}

This subgroup of diffeomorphisms is precisely the conformal group,
$O(4,2)$ and the field $\varphi$ transforms as the dilaton, the Goldstone
boson of a spontaneously broken conformal
symmetry~\cite{Isham:1970gz}. The transformation differs from that of
a Goldstone boson of an internal symmetry crucially in the spacetime
argument of $\tau$. As a consequence, non-derivative interactions, and
in particular a $\varphi^4$ potential is allowed in the low energy
theory.

Requiring non-linearly realized conformal invariance automatically
matches a generally coordinate invariant form of the action, written
in terms of equation~\eqref{eq:metric}, up to the inclusion of a
Wess-Zumino term~\cite{Coradeschi:2013gda},
\begin{align}
  S_{CI}(\tau)
  &=
  S(g)+S_{WZ}(\tau)
  \,.
\end{align}
Therefore, the two theories are identical at the
level of effective field theory. This leads to an important
conclusion: we know what UV completes scalar
gravity (equations~\eqref{eq:weyl} and \eqref{eq:scalar-gravity})! It is a
quantum
CFT on a flat Minkowski background. 

The appearance of a CFT opens up another connection. The AdS/CFT
correspondence allows us to relate the spontaneously broken CFT with
an AdS spacetime, cut off by a IR brane at high
redshifts~\cite{ArkaniHamed:2000ds,Rattazzi:2000hs}. 
The radion is identified with the dilaton in this case. 
Since the radion itself is a geometrical
modulus giving the position of the IR brane,
in this formulation it is not surprising that the curved spacetime of
underlying scalar gravity is the curved spacetime of the IR brane,
described by the radion field. Light matter sees this curved spacetime
by being localized to this brane.

\subsection{Cosmological constant problem in scalar gravity}
\label{sec:cc}

\subsubsection{Classical level}
For general relativity, maximally symmetric solutions exist 
for all values of the cosmological constant, but the Poincar\'{e}
invariant vacuum solution is only obtained at a single fine-tuned 
value of zero CC. This situation is replicated in scalar gravity.

Let us consider the ``chiral Lagrangian'' for scalar
gravity in the Einstein frame on a Minkowski background.
The low energy couplings of the dilaton are fixed by transformations
under
non-linearly realized conformal transformations, and can be simply
written as,
\begin{align}
  \mathcal{L}
  &=
  \frac12 \partial_\mu \varphi \partial^\mu \varphi
  -\lambda \varphi^4
  +\mathcal{L}_{matter}
  \,,
  \label{eq:philaton}
\end{align}
where we have truncated the Lagrangian at two-derivative order, and
$\mathcal{L}_{matter}$ contains ``standard model'' fields. $\varphi$
is coupled to the matter as a ``compensator field'' to ensure
conformal invariance.
We note again that the appearance of a non-derivative coupling for the
Goldstone boson is a consequence of the unique
transformation law shown in equation \eqref{eq:dilaton-transform}.
In fact, the $\lambda$-term  corresponds to 
$\Lambda / M_{pl}^4$ in equation \eqref{eq:scalar-gravity}, the
cosmological constant. 

The presence of the $\lambda\neq 0$ does not admit a
Poincar\'e invariant classical solution. 
We can see that depending upon $\lambda$,
the solutions are~\cite{Coradeschi:2013gda}:
\begin{align}
 \lambda > 0 &\qquad \to \varphi \sim 1/z & SO(3,2)\equiv AdS_4 \,,
 \label{eq:ads4}
 \\
 \lambda < 0 &\qquad \to \varphi \sim 1/t & SO(4,1)\equiv dS_4 \,,
 \label{eq:ds4}
 \\
 \lambda = 0 &\qquad \to \varphi \sim const. & ISO(3,1)\equiv
 \text{Poincar\'{e}}\,.
\end{align}
Generically, the
SO(4,2) group of the CFT is broken to SO(3,2) or to SO(4,1),
corresponding to a classical solution with the dilaton carrying a
non-trivial spacetime dependence.  Matter fields coupled to the
dilaton effectively see an Anti-de Sitter or a de Sitter background.
The form of the dilaton effective potential is fixed by
conformal symmetries, and only admits a translationally invariant
solution upon tuning.
In particular, the hierarchy between the
Planck scale 
and the cosmological
constant in our universe is replicated in scalar gravity for 
$\lambda \sim 10^{-120}$. Strictly speaking, spontaneous conformal
breaking is eliminated for $\lambda \neq 0$, however if $\lambda$ is
small enough it is a useful approximation. For example,
solutions in equations~\eqref{eq:ads4} and~\eqref{eq:ds4} give a
cosmological version of spontaneous breaking.

\subsubsection{Quantum corrections}

In equation \eqref{eq:philaton}, the cosmological constant appears as a marginal
operator. One might then question whether the severity of the fine
tuning here is similar to the cosmological constant
problem in general relativity, where the operator is
super-renormalizable. In principle, since there is no symmetry
protecting the $\varphi^4$ operator, we expect thresholds to generate
the
operator with $\mathcal{O}(1)$ coefficient.
An irreducible
running contribution is generated from the (small) dimensionless
couplings of
the dilaton to the SM
fields, $(v_{weak}/M_{pl})^4$. 
Numerically, we see that the tuning is as severe as in our
universe, where the CC gets irreducible
contributions of at least $v_{weak}^4$.

It turns out that the situation in this case actually parallels the
GR case even more closely. In order to consistently regulate UV
divergences arising from our effective Lagrangian, we need to ensure
that the Ward
identities associated with the conformal invariance are satisfied.
A convenient way to ensure this is to have a $\varphi$-dependent
renormalization scale, 
\begin{align}
  \hat{\mu}\equiv \mu  \frac{\varphi}{M_{pl}}
  \,,
\end{align}
such that vacuum bubbles, two-point functions and four-point functions
(or more generally the Coleman-Weinberg Potential of the dilaton)
appear as 
\begin{align}
  V(\varphi)
  &\sim
  \alpha \hat{\mu}^4 
  +\beta \hat{\mu}^2 \varphi^2
  + \gamma \log[\hat{\mu}^2 / \varphi^2]\varphi^4
  \sim
  \frac{{\mu}^4}{M_{pl}^4} \varphi^4
  \,.
  \label{eq:nocolemanweinberg}
\end{align}
Thus, we see the same matter loop diagrams that contribute to the
spin-2 cosmological constant indeed do contribute to $\lambda$.

\subsubsection{Comparison with the cosmological constant problem in GR}
It is interesting to ask whether the scalar CCP is on the same
footing as the problem in GR. The no-go theorem derived by Weinberg
\cite{Weinberg:1988cp}
focuses on the trace of Einstein's equations, which is of course
identical to the equation of motion for scalar gravity. The crucial
point is that
for translationally invariant solutions, when the matter field satisfy
their equations of motion, general covariance forces the Lagrangian to
have a very specific dependence on $g_{\mu\nu}$,
\begin{align}
  \mathcal{L}
  &= 
  c\sqrt{-g} \,.
\end{align}
Similarly, for scalar gravity, conformal invariance forces the
Lagrangian (again on the classical solutions for matter fields) to be,
\begin{align}
  \mathcal{L}
  &= 
  c \varphi^4 \,.
\end{align}
We see that in either case, there is no non-trivial solution for the
gravitational equation of motion. In order to dynamically relax the
cosmological constant to zero, solutions of matter equations of motion
should imply a solution to the (trace of) Einstein's equations.
Weinberg has argued~\cite{Weinberg:1988cp} that this is not
possible without fine tuning.

Other features of the CCP
\cite{Polchinski:2006gy} which
make a solution hard are also reflected in the scalar gravity case. We
briefly recall some challenges that any solution faces:
\begin{itemize}
  \item Since binding energies and loop corrections to energy
    levels have been measured to
    gravitate, by the equivalence principle these loops should also 
    contribute to
    the cosmological constant.
  \item Modification of gravity at short
    distances
    ($\sim$\SI{100}{\micro\metre})~\cite{Sundrum:1997js,Sundrum:2003jq}
    does not help the situation, since the matter loops in question
    are not cut off at this scale, and the graviton momentum probing
    the CC is
    Hubble scale, nothing to do with the ``compositeness scale'' of
    gravity.
  \item Modification of gravity at very long distances, comparable to
    current Hubble scale runs into the problem that the short distance
    uncancelled cosmological constant prevents the universe from ever
    becoming large enough to probe the very long distance behavior.
  \item Mechanisms which involve gravitational dynamics solving the
    cosmological problem suffer from the problem that the CC only very
    recently became an appreciable contribution to the energy budget,
    so it would be impossible for a mechanism in the early universe to
    operate setting it to be so small.
\end{itemize}
All of these issues apply to the scalar gravity case as well as they
do to spin-2 gravity.
Some of these objections appeal to our cosmic history and some others
merely to the particle physics. We will focus on the particle physics
aspects of the fine tuning. The cosmological mechanisms that address
the issues above are left for future work.

\subsection{Solution: Deformation of the CFT}
We outline the solution here, which arises from 
considering deformations of the CFT~\cite{CPR,Coradeschi:2013gda} (see
also more recent discussions \cite{Chacko:2012sy,
 Chacko:2013dra, Bellazzini:2013fga}).  If we add
a relevant deformation, it explicitly breaks the CFT,
giving a mass to the dilaton.  In order to obtain a regime where the
theory approximates scalar gravity, we would like to have a hierarchy
between the scale of spontaneous CFT breaking (interpreted as the
$M_{pl}$ of scalar gravity) and the mass of the dilaton, the scale at
which there is maximal equivalence principle violation and below which
there
is no long range gravitational force. We would ideally like this
regime to be extremely large to reproduce qualitatively the
exponential hierarchy of scales observed in our universe.

What sets the mass of the dilaton?
The order parameter for CFT breaking is the non-conservation of the
scale current, which is proportional to the $\beta$-function.
For compact internal symmetries, we can usually ensure that the
deformation is small in a controlled fashion, yielding a  light pNGB 
naturally, as is the case for the pion and chiral symmetry in the SM.
However, the present situation for the CFT is different.  The presence
of the
Poincar\'{e}  invariant solution requires that at the scale of
spontaneous breaking, the dilaton potential contributions from the
spontaneous breaking are balanced by those from the explicit breaking.
Thus, we expect the deformation to be $\mathcal{O}(1)$ at the breaking
scale, generically implying an $\mathcal{O}(1)$ $\beta$-function.

The gauge coupling in QCD is an example of such a deformation, and as
the above discussion illustrates, we do not expect a narrow, light
resonance associated with the dilaton. The condition for obtaining a
light dilaton is rather special: the $\beta$-function of
the deformation should stay parametrically small over a range of values of
the coupling. Thus, even when the deformation grows large, the amount
of scale violation is parametrically small. The mass of the dilaton is
suppressed by the small parameter. Since the deformation preserves
the Lorentz subgroup of the CFT, while explicitly breaking scaling,
the low-energy theory still has Poincar\'{e} invariance. 

This dynamical requirement from the CFT point of view is somewhat
mysterious. The AdS dual theory in five dimensions makes the situation
much clearer. The tuning associated with the scalar gravity is nothing
but the tuning of the IR brane tension required in the original
Randall--Sundrum model (RS1 \cite{Randall:1999ee}).  There is of
course an additional tuning of the UV brane itself in that set up,
which is associated with the tuning of the spin-2 cosmological
constant. However, this tuning is decoupled from the IR tuning issue.
In fact, in our analysis, we will always assume that the UV brane is
absent, so
that the dual theory runs to a UV fixed point. Indeed, this is the
theory of scalar gravity, and the spin-2 graviton has been decoupled.

The above discussion suggests inclusion of a Goldberger--Wise
stabilization mechanism in the gravity picture.  The presence of a
small $\beta$-function for the deformation corresponds to a suppressed
bulk potential for the corresponding AdS scalar. Such a suppression
can be protected naturally by an approximate shift symmetry,
in turn realized if the GW scalar is a 5D pNGB of a global symmetry with
a tiny explicit breaking.
While technically natural, we expect all global symmetries
to be at best emergent below the quantum gravity scale.  The fact that
the AdS gravity theory is expected to get strongly coupled not far
from the curvature scale suggests that there may be unacceptably large
violations of the global symmetry by quantum gravity effects.
This is the aspect that we study and control in this paper.

A familiar solution to the problem is to use a gauge symmetry in a
higher dimensions to
obtain the shift symmetry~\cite{ArkaniHamed:2003wu}. A gauge field
in one higher compactified
dimension yields a scalar field in the low energy theory. As a result
of residual gauge symmetry, the scalar possesses a global shift
symmetry which can be robustly protected against quantum effects by
higher-dimensional locality. 
Thus, we will obtain our Goldberger--Wise field in 5D 
from one higher dimension as the sixth component of a
gauge field. The very small potential term is generated by non-local
Aharanov-Bohm phases, which are exponentially suppressed if 6D charged
particle masses are somewhat heavier than the inverse-size of the sixth
dimension. Therefore, we can naturally obtain an
exponentially small potential terms for the GW field.

For fluctuations about the stabilized radion, the (approximate) shift
symmetry for the GW field translates into a shift symmetry for the
dilaton, suppressing its potential.  It is worth noting that the
mechanism involves physics above $M_{pl}$, the scalar
gravity Planck scale, which is also the scale of spontaneous conformal
symmetry breaking. And yet, the mechanism robustly cancels
contributions (from phase transitions or thresholds) far below this
Planck scale.

In sections~\ref{sec:sol4d} and~\ref{sec:sol5d} we review the solution 
originally proposed in
an unpublished work by Contino,
Pomarol, Rattazzi \cite{CPR} and then discussed later in
\cite{Chacko:2012sy, Coradeschi:2013gda, Chacko:2013dra,
Bellazzini:2013fga}. 
The discussion of the mechanism in 4D in section~\ref{sec:sol4d}
highlights the
conditions required for the solution to work, and we 
show that we expect it to be
robust as long as the $\beta$ function stays parametrically small. We
then present a simple example in 5D in section~\ref{sec:sol5d}, where an
approximate shift
symmetry protecting the stabilizing GW field results in the small
$\beta$-function in the 4D effective theory. 
The shift symmetry in 5D is the target for our solution in 6D
presented in section~\ref{sec:sol6d},
where we obtain naturally
the exponentially small potential for the GW field. We show by way of
an explicit computation that all other fields involved in our
calculation decouple and we indeed reproduce the 5D EFT desired. 
We conclude in section \ref{sec:conc}.

\section{The mechanism in four dimensions}
\label{sec:sol4d}

The relaxation mechanism operates
dynamically at low energies in the vacuum such that it robustly
cancels various contributions arising from different scales. This
means that we should be able to study this mechanism purely in terms
of the dilaton effective potential.  We consider the dilaton potential
in the deep IR, after integrating out all matter fields. 

Let us begin by considering the undeformed CFT. Then, the ``SM''
self-couplings, masses and couplings to the dilaton
respect conformal invariance and the IR
dilaton potential is given by,
\begin{align}
  V(\varphi)
  &=
  \lambda \varphi^4
  \,.
\end{align}
This result holds exactly, following from symmetries of the dilaton in 
the non-linearly realized
CFT. In a particular renormalization scheme, care has to be taken in
order for the regulator to not introduce spurious scale dependence in
the potential.
Technically the terminology ``spontaneous breaking of the CFT''
implies that $\varphi$ is a modulus, and hence is only applicable to
the
situation $\lambda=0$. The above equation should be thought of as a
(somewhat small $\lambda$) deviation from this tuned limit. 

We next add a weakly relevant deformation to the CFT, 
\begin{align}
  \mathcal{L}(\mu)
  &=
  \mathcal{L}_{CFT}
  +g(\mu) \mathcal{O}
  \,.
\end{align}
The scale dependence of the coupling constant now affects the
low-energy effective potential.
The $\beta$-function of $g$ is the only source of explicit violation
of the CFT. 
By treating the running coupling as a
spurion~\cite{Chacko:2012sy}, we can write the general form of the
effective potential,
\begin{align}
  V(\varphi)
  &=
  \kappa [g(\varphi)] \varphi^4
  \,,
  \label{eq:veff4d}
\end{align}
where $\kappa$ is a function determined by the strong dynamics of the
CFT.
This Coleman-Weinberg potential generates a mass for the dilaton, and
can stabilize it at a non-zero value.
The stable minimum for this potential can be calculated from
\begin{align}
  \left.
  \frac{\partial V}{\partial \varphi} \right|_{\varphi_{min}}
  &=
  4 \varphi_{min}^3  \kappa [g(\varphi_{min})]
 + 
 \kappa'[g(\varphi_{min}) ] \beta[g(\varphi_{min})] \varphi_{min}^3
  =0
  \label{eq:dvdphi}
  \,.
\end{align}
In order
to balance an $\mathcal{O}(1)$ contribution arising from the spontaneous CFT
breaking, the marginal deformation must itself grow to be
$\mathcal{O}(1)$. At
this point generically conformal invariance is no longer an
approximate symmetry,
since for $\mathcal{O}(1)$ couplings the $\beta$ function itself is not small,
which parametrizes the non-conservation of the scale current,
\begin{align}
  \partial_\mu S^\mu 
  &=
  T_\mu^\mu \propto \beta(g) \,.
\end{align}
We see from above that the generic expectation in the absence of
tuning is that $m_\varphi \sim \varphi_{min}$, which is identified as
the $M_{pl}$
for scalar gravity. Thus, the scalar graviton mass is
expected to be of the order $M_{pl}$, and we do
not obtain a large hierarchy of scales
between which we can approximate the theory as scalar gravity.  This
is analogous to the situation in QCD, where no light dilaton emerges. 

Therefore, we would like to engineer a special situation, where even
when the coupling grows to be $\mathcal{O}(1)$, the $\beta$-function stays
robustly
small. We \emph{assume}
\begin{align}
  \beta(g)
  &=
  \epsilon \bar{\beta}(g)
  \,,
\end{align}
where $\bar{\beta}(g)$ is a generic, $\mathcal{O}(1)$ function of $g$,
with the
only restriction being that it does not have a zero for a finite range
of $g$ (except at $g=0$ where conformal invariance is restored, of course). 

With a slowly varying $\kappa$, we see from equation~\eqref{eq:dvdphi}
that the minimum
of the dilaton potential lies parametrically close to the zero of
$\kappa$. 
At zeroth order in $\epsilon$, $g(\varphi_{min}) = g_*$, where
$\kappa(g_*)=0$. Expanding around this point,
\begin{align}
 g(\varphi_{min})
 &=
 g_*
 -
 \frac{\epsilon}{4}
 \bar{\beta}(g_*)
 +\mathcal{O}(\epsilon^2)
 \,.
\end{align}
We can use this to find $\varphi_{min}$,
\begin{align}
  \varphi_{min}
  &=
  \Lambda
  \exp
  \left[
    \frac{1}{\epsilon}
    \int_{g(\Lambda)}^{g_*-\frac{1}{4}\epsilon \bar{\beta}(g_*)}
  \frac{dg} { \bar{\beta}(g)}
  \right]
  \,,
\end{align}
where $\Lambda$ is a reference scale.
We can expand the potential around the minimum for field fluctuations
$|\delta\varphi|\leq \varphi_{min}$,
\begin{align}
  V(\varphi_{min}+\delta \varphi)
  &=
  \epsilon \frac{\partial \kappa}{\partial g}
  \left(-\frac14\bar{\beta}+\bar{\beta} \log(1+\delta
\varphi/\varphi_{min})
\right)
\left[
  6 \varphi_{min}^2
  \delta \varphi^2
 +\delta \varphi^4
 \right]
 \, .
\end{align}
This shows rather explicitly that the dilaton mass and quartic are
suppressed parametrically over a range of $\delta \varphi$, allowing a
large separation of scale between $m_\varphi$ and $M_{pl}$, as well as
an expanding phase with a tiny scalar cosmological constant if $\varphi$ is
displaced from its potential. 

Note that except for the smallness of the $\beta$ function, we
have made no other assumptions and the
form of the effective potential is fixed by the symmetries. So, we
expect the above conclusions are robust. It is instructive however to
see an explicit example.

Let us first consider the limit $\epsilon=0$, i.e.~the limit of
unbroken CFT, and study an effective Lagrangian for the dilaton
coupling with a fermion,
\begin{align}
  \mathcal{L}_{eff}
  &\ni
  \bar{\psi}\,i \slashed{\partial} \psi
  +y \varphi \bar{\psi}\psi
  +\kappa \varphi^4
  \,.
\end{align}
The fermion gets a mass proportional to its Yukawa coupling, $m_\psi =
\langle \varphi \rangle y$, which is a manifestation of the
equivalence principle. We can study the effective Lagrangian below
this mass scale after integrating out this fermion,
\begin{align}
  \mathcal{L}_{eff}
  &\ni
  \kappa' \varphi^4
  \,.
\end{align}
The form of the potential for the dilaton has unchanged, but the
coefficient gets a correction from the threshold. 
Thus, even if we
started from a zero dilaton quartic just below the scale of CFT
breaking, we would still get contributions from all thresholds below
that scale, similar to what happens for the spin-2 cosmological
constant.
Recall the discussion of equation~\eqref{eq:nocolemanweinberg} that we
do not
get the usual Coleman-Weinberg potential of the form $\sim\varphi^4
\log(\varphi)$, but only $\sim \varphi^4$.

In the presence of the running coupling, the explicit breaking appears
in the effective Lagrangian
as the spurion $g(\mu)$,
\begin{align}
  \mathcal{L}_{eff}
  &=
  \bar{\psi}i \slashed{\partial} \psi
  +y(g(\varphi)) \varphi \bar{\psi}\psi
  +\kappa (g(\varphi)) \varphi^4
  \,.
\end{align}
The spurion $g(\varphi)$ leads to violation of scalar equivalence
principle. At the linearized level, the coupling of $\psi$ to
$\varphi$ is
\begin{align} 
  y(g(\varphi)) +y'(g(\varphi)) \epsilon  \bar{\beta}(g(\varphi))
  \,,
\end{align}
which is no longer tied to the mass of the fermion $\psi$, and we
would observe deviations from the equivalence principle by measuring
the gravitational couplings for multiple $\psi$ species. 
We see that the violation of the equivalence principle is suppressed
by $\epsilon$.
As before, integrating out this fermion results in the
Coleman-Weinberg potential modification of the
function $\kappa$. However, the form of the potential is unchanged
from that in equation~\eqref{eq:veff4d},
with scale
dependence arising solely from $g(\varphi)$, the running of the
coupling near the CFT breaking scale. Thus we see that our conclusions
are robust to matter effects and phase transitions below the CFT
breaking
scale.

We have not yet justified the origin of the assumption, $\beta = \epsilon
\bar{\beta}$. We next show that this can be achieved in a technically
natural way in a variant of an RS model -- a 5D model with an IR brane.

\section{The mechanism in five dimensions}
\label{sec:sol5d}
\label{ssec:5d}
In this section we present a 5D realization of the 4D solution above.
Here we focus on a simple realization in order to focus on the
essential features and look at more general theories in the next
section. The 5D model is a gravitational theory with AdS background,
truncated by an IR brane (that is to say it is the RS1
model~\cite{Randall:1999ee}, but without the UV brane and hence
extending all
the way to $\partial$AdS). We start by outlining the dictionary to
translate
between the AdS and CFT theories.

\subsection{AdS/CFT dictionary}
The AdS/CFT dictionary provides a handy way to identify the
corresponding physics in 5D.
The dilaton is dual to the size of the extra dimension, parametrized by
the radion field,
\begin{align}
  \varphi(x) \leftrightarrow z_{IR}(x)
  \,,
\end{align}
where $z_{IR}$ is the position of the IR brane.
A deformation in the CFT is dual to an AdS scalar field, with the running
coupling identified as (one mode of) the profile of the
scalar, 
\begin{align}
  g(\mu)
  &\leftrightarrow
  \omega (z=\frac{1}{k}\log\mu)
  \,,
\end{align}
where $\omega$ is a scalar field which will play the role of a
Goldberger-Wise stabilizing field~\cite{Goldberger:1999uk}, $k$ is the
AdS curvature and we
have identified the warped
extra dimension coordinate ($z$) as the holographic renormalization scale.
In the limit of a small potential for $\omega$,
the evolution of $\omega$ in $z$ is given by a ``slow-roll''
approximation, such
that $\partial_z^2 \omega \ll k \,\partial_z \omega$. The evolution of the
scalar profile is then simply related to the potential $\partial_z
\omega \simeq \partial V/\partial \omega$, yielding the following
identification,
\begin{align}
  \beta(g(\mu))
  &\leftrightarrow
  \frac{\partial V}{\partial\omega}
  \,.
\end{align}
Let us consider the case where the field $\omega$ is an exact
Nambu-Goldstone boson of a global symmetry, say a $U(1)$. This implies
that there is a shift symmetry for $\omega$, setting $V=0$. This is
dual to a circle of fixed points, where $\beta=0$. This is expected
from the fact that the spontaneous breaking of $U(1)$ leads to a set
of degenerate AdS vacua, each of which corresponds to a CFT. Since
$U(1)$ SSB is robust in AdS effective field theory, an approximate
shift symmetry for $\omega$ can also be realized robustly, leading to
a small potential, and hence a small $\beta$ function\footnote[2]{In
  AdS quantum gravity some breaking of the $U(1)$ global symmetry --
  and
  hence the shift symmetry of $\omega$ -- is to be expected. Indeed,
  it is to control this aspect of the problem that we present a 6D
  construction in section~\ref{sec:sol6d}. 
  Here we take the AdS
  effective field theory view that small explicit breaking is natural.
}.
In order to obtain a slightly relevant deformation we can add a small
negative mass-squared for the scalar (which is stable on an AdS
background).  Once we turn on the small potential and source it at the
$\partial$AdS, the scalar field backreacts on the metric taking it
away from the pure AdS limit. This corresponds to a
breaking of the CFT via running of the dual coupling.  

The ``SM matter'' terms in the dilaton effective action
(equation~\eqref{eq:philaton}) appear as
brane localized terms in the 5D picture,
\begin{align}
  \mathcal{L}_{matter}
  &\leftrightarrow
  \mathcal{L}_{brane,IR}
  \,.
\end{align}
The cosmological constant problem in scalar gravity is dual
to the IR brane tension tuning in RS1 model,
\begin{align}
  \lambda \varphi^4
  &\leftrightarrow
  \left[\sqrt{g}\, \delta T \right]_{IR}
  \,,
\end{align}
where $\delta T_{IR}$ is the detuning of the brane tension from the
value in RS1 required to tune the radion potential to zero. It
includes the vacuum energy contributions from the SM fields living on
the IR brane. Upon including the deformation, the correspondence
becomes 
\begin{align}
  \kappa(g(\varphi)) \varphi^4
  &\leftrightarrow
  \left[\sqrt{g} \,f(\omega)\right]_{IR}
  \,,
\end{align}
where we have combined the brane tension detuning and couplings to
$\omega$ into a single function
$f(\omega)$. 
We see that we want the Goldberger-Wise scalar to have a 
generic potential localized on the IR brane. Locality preserves the
approximate global symmetry in the bulk even though it is broken badly
on the IR brane. 
We discuss a
simple realization of these features next.

\subsection{5D classical solution}
We review the simple version of a 5D model presented in
\cite{Coradeschi:2013gda}. For computational simplicity we work in the
case where backreaction of
the field is somewhat small everywhere. As shown in
\cite{Coradeschi:2013gda}, this
assumption is not needed. 
Since
we are tracking a fine-tuning of O(10$^{-120}$), taking some
parameters to be small (but $\mathcal{O}(1)$) to maintain perturbative
control should be harmless.

The model is,
\begin{align}
  \frac{S}{M_5^3}
&=
\int d^4x \, dz\,
\sqrt{G}
\left[
-\frac14 R 
+ 3k^2 
+\frac12 (\partial \omega)^2+2 k^2 \epsilon\, \omega^2
\right]
-\frac12
\int_{IR} d^4x
\sqrt{g}\,
k \left[-3 +
 f(\omega)
\right]
  \,,
\end{align}
where $\omega$ is a dimensionless, pNGB field 
(which is denoted $\pi$ in \cite{Coradeschi:2013gda}),
and $f(\omega)$ is a generic brane-localised potential for $\omega$.
Note that we are working in the $(+---\ldots)$ signature. 

The model above is clearly not the most general Lagrangian consistent
with the approximate
shift symmetry of the Goldstone. However, it serves to illustrate the
key features of the solution, and we consider robustness against 
generalizations, other
deformations and quantum corrections later.

We first seek a 4D Poincar\'{e} invariant
ground state within the domain of the EFT,
\begin{align}
  ds^2 &= e^{2A(z)} dx^\mu dx^\nu \eta_{\mu\nu} - dz^2 
  \,,
  \\
  \omega (x,z) &= \omega(z)
  \,.
\end{align}
The equation of motions for $A, \omega$ in the bulk are,
\begin{align}
  A'^2 - k^2 -\frac23 k^2 \epsilon\, \omega^2- \frac16 \omega'^2 
  &=0 
  \,,
  \\
  \omega'' + 4 A' \omega' + 4 k^2 \epsilon\, \omega
  &=0
  \,,
  \\
  A''+ \frac23 \omega'^2 
  &= 0
  \,,
\end{align}
and the junction matching conditions on the IR brane are,
\begin{align}
  \omega'(z = z_{IR} ) 
  &= -\frac12 k \,\frac{\partial f(\omega_{IR})}{\partial \omega}
  \,,
  \\
  A'(z=z_{IR})
  &= - k  \left[1 + \frac13  f(\omega_{IR}) \right]
  \,.
\end{align}

If we neglect the backreaction of the 
GW field on the metric, then we can write the solution for its
equation of motion in a background AdS space 
\begin{align}
  \omega 
  &= 
  \omega_* \, e^{\Delta_- k z} 
  + \hat{\omega}\, e^{\Delta_+ k (z-z_{IR})}
  \,,
\end{align}
where $\Delta_\pm = 2(1\pm\sqrt{1-\epsilon})$ .  This is the familiar
Goldberger-Wise scalar~\cite{Goldberger:1999uk} profile used to
stabilize the Randall-Sundrum branes. 

When is our assumption of small backreaction justified? Since the
potential $V(\omega)\sim \mathcal{O}(\epsilon)$, the dominant
backreaction comes
from the kinetic term for $\omega$.
The contribution from the slowly varying $e^{\Delta_- kz}$ term is
small, so that the backreaction is determined by the size of
$e^{\Delta_+k(z-z_{IR})}$ term. To ensure that this is small even as
we get close
to IR brane, we need $\hat{\omega}$ to be
parametrically small. 
Let us first consider if we can have
$\hat{\omega}=\mathcal{O}(\epsilon)$. The junction conditions are then
only satisfied if $f(\omega_{IR})={\partial f(\omega_{IR})}/{\partial
\omega}=\mathcal{O}(\epsilon)$, which reintroduces the
tuning of the brane tension on the IR brane.
Let us allow a detuning of the brane tension away from this limit,
with the detuning set by
a moderately small parameter which we will call $\eta$. 
In this case the function $f(\omega)$ is chosen such that there
exists a value $\omega=\bar{\omega}$ such that
$f(\bar{\omega})\sim\partial f(\bar{\omega})/\partial \omega \sim
\mathcal{O}(\eta)$. This corresponds to a mild tuning of
the IR brane tension. We can treat the backreaction perturbatively in
$\eta$.

With this mild tuning, the boundary matching conditions are satisfied
by our solutions at 
zeroth order in detuning parameter $\eta$.
The IR brane position is fixed at 
\begin{align}
  z_{IR}
  &\simeq
  \frac{1}{k \Delta_-} 
  \log \left[ \frac{\omega_*}{\bar{\omega}} \right]
  \,.
\end{align}
This suggests that the radion has been stabilized.
At higher orders in $\eta$, $\hat{\omega}$ is non-zero, and the GW
profile
backreacts on the metric. The solution can be self-consistently
solved for order by order in $\eta$. We present this solution
later in section~\ref{sec:sol6d}, and now turn to the effective
potential for
the radion and the mass of the radion fluctuations.

\subsection{Effective action for the radion}
It is instructive to derive the 4D effective action for the
radion $r(x)$ by plugging in the solution above back into the 5D
action. The metric with the radion fluctuation is conveniently
parametrized as~\cite{Charmousis:1999rg},
\begin{align}
  ds^2 &= e^{-2 k (z + r(x)e^{2 k z})} dx^\mu dx^\nu \eta_{\mu\nu} 
  - (1+2 k r(x) e^{2 k z}) dz^2 
  \,.
\end{align}

The difference relative to the RS case is the behavior of the
fields towards the
AdS boundary (which is cut off by the UV brane in the usual
Randall-Sundrum models). In the present case, we need to add a
``regulator'', for which we introduce a
boundary at $z_{UV}$. The presence of the boundary yields
a finite action upon dimensional reduction, and the boundary terms
are also required for a well-defined variational
principle~\cite{Polchinski:2010hw}. The boundary term is
\begin{align}
  S
  &\supset
  -\frac{M_5^3}{2}
  \int_{UV} d^4 x \sqrt{g}
  \left[3k + \Delta_- \, k\, \omega^2 \right]
  \, .
\end{align}
This is analogous to UV regulating the 4D CFT. This boundary term
ensures that our deformation is a free input, parametrized by
$\omega_*$, the co-efficient of the near-boundary behavior of the
scalar field. It also ensures that the near-boundary geometry is pure
AdS.

The kinetic piece of the radion action is
given by,
\begin{align}
  \mathcal{L}_{kin}
  &=
  \frac34 k M_5^3 \left(
  \exp\left[{2k  (z_{IR}-e^{2k z_{IR}} r(x))}\right]
  -\exp\left[{2k  (z_{UV}-e^{2k z_{UV}} r(x))}\right]
  \right)
  \partial_\mu r \partial^\mu r
  \,.
\end{align}
For the calculation of the potential, it is sufficient to look at the 
limit where the fluctuation $r(x)=0$. In the bulk, 
\begin{align}
  -\frac{V_{bulk}}{M_5^3}
&=
\frac k2
e^{-4 k z_{IR}}
\left[
  (1-e^{-4 k  (z_{UV}-z_{IR})})
  -\Delta_+\hat{\omega}^2
  (1
  - e^{(\Delta_+ - \Delta_-) k (z_{UV}-z_{IR})})
\right]
\nonumber\\& \qquad\qquad\qquad
-\frac{ k \Delta_- } {2}
\omega_*^2 
( e^{-(\Delta_+ - \Delta_-) k z_{IR}}
- e^{-(\Delta_+ - \Delta_-) k z_{UV}})
\,.
\end{align}
We see that there are bulk contributions which diverge as
$z_{UV}\to-\infty$.
There is an extra contribution from the extrinsic
curvature of the UV and IR branes,
\begin{align}
  R|_{|z_{IR}|} &= 8 k \delta(z-z_{IR})
  \,, \quad
  R|_{|z_{UV}|} = -8 k \delta(z-z_{UV})
  \,.
\end{align}
If the branes are thought of as orbifold fixed points, this
contribution arises as the contribution of the kink to
the curvature. Thus, the brane contributions to the potential are,
\begin{align}
  -\frac{V_{brane,IR}}{M_5^3}
&=
-\frac12
e^{-4 k z_{IR}}
\left[k +
  k f(\omega(z_{IR}))
\right]
\,,
\\
-\frac{V_{brane,UV}}{M_5^3}
&=
-\frac12
e^{-4 k  z_{UV}}
\left[-k +
  k \Delta_- \omega(z_{UV})^2
\right]
\,.
\end{align}

The total potential is given by
\begin{align}
  -\frac{V}{M_5^3}
  &=
  -\frac{k}{2 }
e^{-4 k  z_{IR}}
  \left[
   \Delta_+\hat{\omega}^2
  (1
  - e^{(\Delta_+-\Delta_-) k (z_{UV}-z_{IR})})
  + f(\omega(z_{IR}))
\right]
  \nonumber \\& \qquad\qquad\qquad
  -\frac{ k \Delta_- } {2}
\left[
  \omega_*^2 
  e^{(\Delta_- - \Delta_+) k \, z_{IR}}
  +
    2 \omega_* \,  \hat{\omega}\, 
e^{-\Delta_+ k  z_{IR}}
    + \hat{\omega}\,^2 e^{(\Delta_+-\Delta_-) k \, z_{UV}}
e^{-2 \Delta_+  z_{IR}}
\right]
  \,.
\end{align}
We see that the regulator cancels the potentially dangerous 
terms which would blow up as $z_{UV} \to \infty$, and so we can
proceed to that limit,
\begin{align}
  \frac{V}{M_5^3}
  &=
  \frac{k }{2 }
e^{-4 k  z_{IR}}
\left(
  \Delta_+ \hat{\omega}^2
  +\ f(\omega(z_{IR}))
  \right)
  +\frac{ k \Delta_- } {2}
e^{-\Delta_+ k  z_{IR}}
  \left ( 
  \omega_*^2 e^{\Delta_- kz_{IR}}
  +2 \omega_* \,  \hat{\omega}\, 
  \right)
  \,.
\end{align}
Matching conditions are given by,
\begin{align}
  \omega'(z_{IR})
  &=
  -\frac12
  \frac{\partial f(\omega(z_{IR}))}{\partial \omega }
  \,,
\end{align}
implying,
\begin{align}
  \hat{\omega} 
  &=
  -\frac{1}{2 k \Delta_+} \frac{\partial f(\omega_{IR})}{ \partial \omega}
  - \frac{\Delta_-}{\Delta_+} \omega_* e^{\Delta_- k z_{IR}}
  \,.
\end{align}
This also implies that 
\begin{align}
  \omega(z_{IR})
  &=
  -\frac{1}{2 k  \Delta_+} \frac{\partial f}{ \partial \omega}
  - \frac{\Delta_+-\Delta_-}{\Delta_+} 
  \omega_* e^{\Delta_- k z_{IR}}
  \equiv 
  \sigma(
  \omega_* e^{\Delta_- k z_{IR}}
  )
  \,.
\end{align}
Identifying the canonical radion,
\begin{align}
  \varphi(x)
  &= 
  f e^{-k (z_{IR}+ r(x) \exp({2 k z_{IR}))}}
  \,,
\end{align}
where $f^2=\frac32 M_5^3 / k$, we get the canonically normalized
kinetic term for $\varphi$ from $\mathcal{L}_{kin}$. The potential can
be calculated by taking the fluctuation $r(x)=0$, as before. We also
identify $\omega_* e^{\Delta_- k z_{IR}} =
g(\varphi)$ as the running coupling, which is weakly varying
($\sim \varphi^{-\epsilon}$).
Thus the effective Lagrangian (ignoring terms suppressed by powers of
$\epsilon$,
\begin{align}
  \mathcal{L}
  &=
  \frac12 \partial_\mu \varphi \partial^\mu \varphi
  -\frac {M_5^3 k} {2 f^4} \varphi^4
  \left[
  \frac{1}{16}
  \left[
    \frac{\partial f(\sigma(g(\varphi)))}{\partial g}
  \right]^2
  + f(\sigma(g(\varphi)))
\right]
\,.
\end{align}
This matches the result of~\cite{Coradeschi:2013gda} for small
backreaction. The term in the square brackets is a function
$\kappa(\varphi)$, which is slowly varying by virtue of the running
coupling being near marginal. Thus it is of the desired form
in equation~\eqref{eq:veff4d}.

We have only considered a mass term for the $\omega$ field, and adding
other potential terms makes it challenging to
obtain an analytical solution. The higher order terms in the $\omega$
potential are subdominant if we additionally assume $\omega_*$ to be
small. That is, we assume that the deformation stabilizes CFT breaking
at weak coupling. In this case, we can
safely ignore higher order terms in the potential for $\omega$, and
only keep the $\epsilon k^2$ mass term.
However, as we will see in the next section this approximation can be
relaxed while preserving the qualitative mechanism.

The assumptions made in this section highlight the checks we need to
perform. Since we have been working to leading order in $\eta$, we
need to ensure that higher order terms in $\eta$ do not spoil our
mechanism.  Crucially, we need the global symmetry of the $\omega$
field to be only broken by $\epsilon$, even at the level of
non-renormalizable interactions in 5D. This issue certainly needs
addressing in presence of a quantum-gravitational threshold which is
expected to generate shift-symmetry violating corrections to the
$\omega$ potential. From the AdS/CFT point of view, it is not clear
what this global symmetry in the bulk corresponds to in the CFT. We
have also not yet included other higher dimensional operators
consistent with the shift symmetry, or studied carefully the effect of
quantum corrections. We address these issues next in our 6D
construction, where $\omega$ is identified as the sixth component of
6D gauge field.

\section{Six dimensions : Goldberger-Wise field as $A_6$}
\label{sec:sol6d}
The solution outlined in the previous section sets the stage for the
6D solution. In the 6D effective field theory, all aspects of our
mechanism are robustly treated in the infrared, insensitive to further
UV completion.  We first consider an action in 6D, and write down an
explicit classical solution. This serves to identify the heavy fields
which can be integrated out in the 5D effective theory.  As mentioned
before, once we ensure that the shift symmetry for Goldberger-Wise
field is of high quality, subleading corrections to the simple
solution above will not affect the basic mechanism.  Our goal in this
section will be to show that we obtain the simple 5D theory presented
above with only subleading corrections, while focusing on protecting
the approximate shift symmetry for GW.

\subsection{Action}
We parametrize the action in six dimensions as
\begin{align}
  \frac{S}{M_6^4}
  &= 
  \int d^4x \, dz\, d\vartheta \, 
  \sqrt{G}
  \left[
    -\frac14 R
    + 4 k_5^2
  +\mathcal{L}_{axion}
  +\mathcal{L}_{gauge}
\right]
\nonumber \\&\qquad
  + \frac12
  \int d^4x \, dz\, d\vartheta \, 
  \sqrt{\bar{g}}
  \delta(z-z_{IR}) 
  \,
  \mathcal{L}_{brane}
  \,.
\end{align}
The 6D radius is stabilized by  an axion wrapped around the sixth
dimension,
\begin{align}
  \mathcal{L}_{axion}
  &=
  \partial_a \chi^\dagger \partial_b \chi G^{ab}
  \,.
\end{align}
We treat the axion in a non-linear sigma model, $\chi = v e^{i\sigma}$.
The value $v$ on the brane can differ from the one in bulk in general
without affecting our argument.
Note that while our simple Lagrangian for the axion respects a global
$U(1)$ symmetry, this symmetry is not crucial for our solution and is
merely retained for simplicity. In particular, there are no light 5D
fields associated with this symmetry (appendix~\ref{sec:zeromodes}),
and hence $U(1)$ symmetry breaking deformations do not affect the form
of the 5D effective action we obtain.
The 6D bulk contains a gauge field and a heavy charged
scalar
\begin{align}
  \mathcal{L}_{gauge}
  &=
  -\frac{1}{4} F_{mn} F_{ab} G^{ma} G^{nb}
  +D_a \xi^\dagger D_b \xi G^{ab}
- m_\xi^2  \xi^\dagger \xi
\,.
\end{align}
where $D_a = \partial_a - i e A_a$. Note that the gauge coupling has
dimension $[e]=1$ due to our choice of normalization.
The  non-local Wilson loop along a curve $\gamma$ that winds around the
compactified sixth dimension, $\oint_{\gamma} e  A_6$,
matches on to the Goldberger--Wise scalar in 5D. Specifically we
identify, \begin{align}
  \omega(z) 
  &=  
  \frac{1}{2\pi \rho(z) e}\oint_{\gamma} e A_6
  = A_6(z)
  \,.
\end{align}
Here, $\rho$ is the radius of the sixth-dimension which has
$z$-dependence in general, and 
we are working in the ``almost-axial'' gauge where $A_6$ does not
depend on $\vartheta$.
This Aharanov-Bohm phase can only be detected by loops of
charged matter that wind around the circle. 
Therefore, the leading dependence in the 6D effective action on 
the Wilson loop will arise from such loops of the lightest charged
particles, and will match on the 5D GW potential for $\omega$.
In the bulk, the only charged field
is a heavy (compared to the 6D KK scale) scalar $\xi$, and its
loop contribution is suppressed exponentially (see appendix
\ref{app:casimir}),
\begin{align}
  V_{bulk}(\rho,\omega)
  &
  \simeq
  \epsilon M_6^2 \cos (2\pi  e \omega)
  \,,
  \label{eq:vbulk}
\end{align}
where the small parameter $\epsilon$ is defined as
\begin{align}
  \epsilon
  &=
  \frac{1}{2\pi\rho M_6^6}
  \left(\frac{m_\xi}{4\pi^2 \rho} \right)^{5/2}
  e^{-2\pi \rho m_\xi }
  \,.
\end{align}

The action on the IR brane is assumed to contain light fields that
appear as SM matter fields in the 4D effective theory. It also
contains some light fields charged under the gauge group. The exact
nature of the lights fields will not matter, only that they give rise
to a generic brane potential for the sixth component, $A_6$, of the
gauge field.  Assuming that these light fields have masses
$m\lesssim1/\rho(z_{IR})$,  this potential is unsuppressed.  In
general, for multiple light fields with different $U(1)$ charges and
masses we get a sum of terms, and thus can obtain a generic dependence
on $\omega$ at the brane (see appendix \ref{app:casimir}).  The brane
potential is given by
\begin{align}
  V_{brane}(\rho,\omega)
  &=
  \tau
  -\frac{\sqrt{2}}{3(2\pi)^3\rho^5 M_6^4}
  \cos(2\pi e \omega)
  +\eta f(v,\rho)
  \,.
\end{align}

We note that at this stage upon the 6D compactification we also obtain
a 5D gauge
field, $A_M$. 
However, a general boundary condition on the brane at $z_{IR}$ 
ensures that $A_M$ zero mode does not survive down to the low-energy
4D effective theory.

\subsection{Equations of motion}

We first assume circular symmetry in the $S_1$ direction and
4D Poincar\'{e} invariance. The  ansatz parametrizing this
is,
\begin{align}
  ds^2
  &=
  e^{2A(z)} \eta_{\mu\nu} dx^\mu dx^\nu
  -dz^2
  -\rho(z)^2 d\vartheta^2 ,
  \\
  \chi &= v e^{i \vartheta},
  \quad
  \xi =0, \quad
  A_M =0, \quad
  A_6 = \omega(z) 
  \,.
\end{align}
The gravity equations of motion (after some algebra) read 
\begin{align}
  -\frac52 A'^2 - A'' 
  +\frac14  \frac{\omega'^2}{\rho^2}
  +\frac{v^2}{2\rho^2} 
  -\frac12 
    \rho \frac{\partial V}{\partial
  \rho}
  +\frac12 V(\omega,\rho)
 & =\frac14
  \delta(z-z_{IR})
  \left[
    V_{brane}
    +\rho \frac{\partial V_{brane}}{\partial \rho}
  \right]
  \,,
  \label{eq:6DGR1}
  \\
  -\frac32 A'^2 -\frac{A' \rho'}{\rho}
  +\frac{1}{4} \frac{\omega'^2 }{\rho^2}
  -\frac{v^2}{2\rho^2} 
  +\frac12 V(\omega,\rho)
  &=0
  \label{eq:6DGR2}
  \,,
  \\
  \frac{\rho''}{\rho}
  +\frac{4 A'\rho'}{\rho} 
  +\frac32 \frac{\omega'^2}{\rho^2}
  +\frac{4v^2}{\rho^2}
  -\frac{3\rho}{2} \frac{\partial V}{\partial \rho}
  -V(\omega,\rho)
  &=
    -\frac{1}{4} 
  \delta(z-z_{IR})
  \left[
    V_{brane}
    -3\rho \frac{\partial V_{brane}}{\partial \rho}
  \right]
  \,,
  \label{eq:6DGR3}
\end{align}
where for brevity we have redefined,
\begin{align}
  \left(
    4k_5^2 
    -V_{bulk}(\omega,\rho)
    \right)
    &= V (\omega,\rho)
    \,.
\end{align}
The equation of motion for $\omega(z)$,
\begin{align}
  \omega''+ 4 A'\omega'
  -\frac{\omega'\rho'}{\rho}
  -\rho^2 \frac{\partial V}{\partial \omega}
  &=
  \frac{\rho^2}{2}
  \delta(z-z_{IR})
  \frac{\partial V_{brane}}{\partial \omega}
  \,.
\end{align}

\subsection{Solution at $\epsilon=0$} 
We first work in the limit
$V_{bulk}(\omega,\rho)=0$. This is the limit of exact scalar gravity, so we
will find a massless scalar graviton solution after
tuning the radion potential to zero. We work perturbatively in
the
backreaction, parametrized by $\eta$. We parametrize the solutions as,
\begin{align}
  \rho(z) &= \bar{\rho} + \eta \rho_1 (z) +\ldots\\
  \omega(z) &= \bar{\omega} + \eta \omega_1 (z) +\ldots\\
  A'(z) &=  -k_5+ \eta A'_1 (z) +\ldots
\end{align}

The limit of negligible backreaction requires the brane terms to be
tuned such that the geometry does not deviate from AdS all the way to
the IR
brane. Thus, there exists a solution 
with $\omega(z), \rho(z) = const$. 
Matching this solution at the boundary will provide us
with the fine tuning we need to perform at zeroth order in $\eta$. 
This ansatz leads to the following equations of motion in the bulk,
\begin{align}
  -\frac32 A'^2 + \frac12 V(\bar{\omega},\bar{\rho}) -
  \frac{v^2}{2\bar{\rho} ^2}
  &=0
  \,,
  \\
  -\frac52 A'^2 
  +\frac12 V(\bar{\omega},\bar{\rho}) 
  +\frac{v^2}{2\bar{\rho} ^2} 
  &=0
  \,,
  \\
  - V(\bar{\omega},\bar{\rho} )
  +\frac{4v^2}{\bar{\rho} ^2}
  &=
  0
  \,.
\end{align}
The $\omega(z)$ equation is satisfied trivially for $\epsilon=0$ for any
constant $\omega(z)$.

The solution for $A'$ and $\bar{\rho} $ is,
\begin{align}
  A'^2
  &= k_5^2 \,,\\
  \frac{v^2}{\bar{\rho}^2} 
  &=
  k_5^2
  \,.
\end{align}
We see that this results in an AdS$_5$ space, and the 6D radius has
been stabilized. 
The boundary conditions are given by,
\begin{align}
  \frac{\partial V_{brane} (\bar{\omega},\bar{\rho} ) }
  {\partial \omega}
  &=0
  \,,
  \\
  V_{brane}(\bar{\omega},\bar{\rho} ) 
  + \bar{\rho}  
  \frac{\partial V_{brane}}{\partial \rho} 
  &=
  -4k_5 
  \,,
  \\
  V_{brane}(\bar{\omega},\bar{\rho} ) 
  - 3 \bar{\rho}  \frac{\partial V_{brane}}{\partial \rho}
  &=
  0
  \,.
\end{align}
Since the bulk solution already fixes all integration constants, we
see that the boundary conditions are satisfied to
this order by a
tuning of $\mathcal{O}(\eta)$.
The brane potential terms need to
be tuned in order to be consistent with the ansatz that $\omega(z)$
and $\rho(z)$ are constant. 
The fact that the asymptotic value of the deformation, $\bar{\omega}$,
is
fixed by the junction matching conditions, will persist at
($\epsilon=0$) at every order in $\eta$.

Let us now go on to first order in $\eta$. 
For the first order terms, the GR equations in the bulk are
\begin{align}
  \frac{\rho_1''}{\bar{\rho} } - 4 k_5 \frac{\rho_1' }{\bar{\rho} }
  - \frac{4v^2}{\bar{\rho} ^3} \rho_1 
  &=0 
  \,,
  \\
  - A_1''(z)
  +5  k_5 A_1'(z) 
  -\frac{v^2}{\bar{\rho} ^3}\rho_1
  &=0
  \,,
  \\
  \label{eq:A1firstorder}
  k_5 \frac{\rho_1'}{\bar{\rho} } + 3 k_5 A_1' 
  +\frac{v^2}{\bar{\rho} ^3}\rho_1
  &=0
  \,.
\end{align}
The matter equation of motion for $\omega$ is,
\begin{align}
\omega_1'' - 4 k_5 \omega_1' &= 0 
\,.
\end{align}
As usual, there is some redundancy in these equations. In particular,
the overall constant in $A_1(z)$ (say $A_1(0)$) is unphysical, and we
set it to zero.
Equation \eqref{eq:A1firstorder} shows that $A_1'(z)$ is determined
algebraically once $\omega_1(z)$ and $\rho_1(z)$ are fixed. Thus, there
are four unknown constants of integration, two each associated with
second order differential equations for $\rho_1(z)$ and
$\omega_1(z)$.

The boundary conditions at this order are given by,
\begin{align}
  -\eta \omega_1'(z_{IR})
  &=
  \frac{\bar{\rho} ^2}{2} \frac{\partial V_{brane}}{\partial \omega}
  \,,
  \\
  \eta A_1'(z_{IR})
  &=
  \frac14 
  \left[
    V_{brane}+\rho \frac{\partial V_{brane}}{\partial \rho}
  \right]
  \,,
  \\
  - \eta\rho_1'(z_{IR}) 
  &=
  -\frac14 
  \left[
    V_{brane}-3\rho \frac{\partial V_{brane}}{\partial \rho}
  \right]
  \,.
\end{align}
Recall that the brane terms were tuned to $\mathcal{O}(\eta)$, so
they appear as generic
$\mathcal{O}(1)$ brane terms at this order, fixing 3 integration
constants.
The other undetermined constant is fixed by requiring
a finite energy solution, which implies that
$\rho(z)|_{z\to-\infty}=const$.

The solutions are of the form, 
\begin{align}
  \rho_1(z) &= c_1 e^{k_5\Delta_\rho (z-z_{IR})}\,,\\
  \omega_1(z)  &= c_2 +  c_3 e^{4 k_5 (z-z_{IR})}\,,\\
  A_1'(z)  &= -c_1 \frac{k_5 (\Delta_\rho+1)}{3\bar{\rho} }
  e^{\Delta_\rho (z-z_{IR})}
  \,,
\end{align}
where $\Delta_\rho \simeq 2(1+\sqrt{2})$ is the scaling dimension
of the operator corresponding to the $\rho$ deformation. The three constants
$c_i$ are fixed by the three junction matching conditions. Higher order 
terms in $\eta$ all have a similar
exponential behavior turning on near the IR boundary.
From the holographic dictionary it is clear that the $z$ dependence of 
$\rho$ and $\omega$ is dual to these deformations picking up vevs
after spontaneous symmetry breaking. Note that currently the IR brane
position
($z_{IR}$) is not fixed, since a translation of the brane does not
change the asymptotic solution as $z\to-\infty$. This is again
consistent with the fact that we only have a spontaneous breaking of
the CFT, leading to no potential for the dilaton.

We get a Poincar\'{e} invariant solution, so what happened to the
quartic term in the dilaton potential?
We see that all the undetermined constants are fixed by the boundary
conditions. In particular the boundary value,
$\omega_1(z)|_{z\to-\infty}$ is fixed by the IR boundary conditions.
Thus, at every order in $\eta$, we need to perform one tuning of
the boundary value of $\omega$ in order to obtain a Poincar\'{e}
invariant ground state.

\subsection{Turning on $\omega$ potential in the bulk}
We now take the quantum correction to the potential of $\omega$ into
account generated by the loops of charged matter as in
equation~\eqref{eq:vbulk}.

Note that the fine tuning in the $\epsilon=0$ solution above 
arises because the IR boundary condition
essentially fixes the UV boundary condition (at the AdS boundary, or
any other convenient $z\ll 0$) for $\omega$. Once we introduce a
potential for $\omega$ in the bulk, there is a slowly varying profile
for $\omega$ in the bulk, effectively scanning over different values
of $\omega(z)$ with varying $z$. Thus, we expect the modulus to be
stabilized close to where the value of $\omega(z)$
approaches $\bar{\omega}$.

Unlike the 5D example in section \ref{ssec:5d}, an analytical solution
is generically not feasible to obtain, even for small backreaction.
However, an approximate solution can be obtained using singular
perturbation theory
\cite{Chacko:2013dra,Bellazzini:2013fga,Coradeschi:2013gda}.  
There are two separate
qualitative regions, which can be matched at $\bar{z} \sim z_{IR}-\log
\epsilon$. For $z > \bar{z}$, close to the IR brane, the $\epsilon$
perturbation is subdominant to the contribution from derivatives of
$\omega, \rho, \pi, A$ etc.  Thus, the solution found above applies at
leading order in $\epsilon$.  In particular, this solution requires
that $\omega (z) = \bar{\omega} + \mathcal{O}(\epsilon)$ for $z\sim
\bar{z}$.

The $\epsilon=0$ solutions enter their asymptotic behavior for $z
\lesssim \bar{z}$. In this asymptotic region, the backreaction
from $\omega$ on the metric is small. In fact, in this region we can
robustly say,
\begin{align}
  \rho(z) &= \bar{\rho} +\mathcal{O}(\epsilon) \,,\\
  A'(z) &= -k_5 +\mathcal{O}(\epsilon) 
  \,.
\end{align}
Additionally, the variation in $z$ is controlled by the potential, 
so that each derivative
is suppressed by additional powers of $\epsilon$. Thus, the leading
order effect of $\epsilon$ for these fields is to change their
matching condition at $z=\bar{z}$ by $\epsilon$.

The dominant effect is of course on the profile of $\omega$ itself,
where it can now slowly evolve to zero at the AdS boundary.
Inspecting the
equations of motion, we see that to leading order in $\epsilon$, the
solution in this region is given by a first order differential
equation for $\omega$.
\begin{align}
  \omega'
  &=
  -\frac{\bar{\rho}^2}{4k_5}\frac{\partial V}{\partial \omega}
  =
  \frac{2 \pi e\, \epsilon\,}{k_5} \bar{\rho}^2 M_6^2 \sin(2\pi e\omega)
  \,.
\end{align}
The above equation is easily solved as is. For the sake of making
contact with the discussion in section~\ref{ssec:5d}, let us consider the
limit $2 \pi e\omega
\ll 1$, so that
we can write,
\begin{align}
\omega'
&\simeq
\frac{4 \pi^2 e^2}{k_5} \epsilon \bar{\rho}^2 M_6^2 \, \omega (z)
\,,
\quad
\Rightarrow
\omega(z)
=
\omega_*
e^{\frac{4 \pi^2 e^2}{k_5} \bar{\rho}^2 M_6^2\epsilon z}
\,.
\end{align}

The IR brane requires a specific value  for the
asymptotic
value of $\omega$ (say $\omega(\bar{z})=\bar{\omega}$). We match our
solution to this value at $\bar{z}$.
The matching condition yields,
\begin{align}
  \bar{z}
  =
  \frac{k_5}{4 \pi^2 e^2 \bar{\rho}^2 M_6^2\epsilon }
  \log \frac{\omega_*}{\bar{\omega}}
  \,.
\end{align}
which leads to the familiar result we obtained above,
 $\bar{z} \sim \frac{1}{\epsilon}$ . (Note that $z=0$
corresponds to the
reference scale
where  our
deformation coupling $\omega_*$ is defined). The brane is stabilized
at $z_{IR} \sim \bar{z} + \log\epsilon$. We see that we have now
traded the parameter $\omega_*$ for $z_{IR}$; we are free to choose
any asymptotic value for the deformation (defined at a reference scale),
and that fixes the value of $z_{IR}$, the location where the IR brane
is stabilized.

The only light fields in the bulk we have are $\omega$  (see
appendix~\ref{sec:zeromodes}) and $A_M$, in addition to light fields
on the
brane. Therefore, we can move to a 5D effective description,
integrating out physics above the scale $1/\bar{\rho} $.

\subsection{General solution in 5D}

In order to make connection with the 5D example earlier, we
dimensionally reduce our 6D theory to 5D. The scale $1/\bar{\rho} $
serves
as the heavy threshold. The
low energy degrees of freedom are the pseudo-Nambu Goldstone,
$\omega$, the $U(1)$ gauge field $A_M$ (which will not play a role
here) and other light matter fields on the IR brane. Therefore,
the 5D effective Lagrangian looks like,
\begin{align}
  \mathcal{L}
  &=
  M_5^3\int d^4 x dz \,
  \sqrt{G}
  \left[
  -\frac14 R
  + 3k_{new}^2
  +\frac12 \partial_a \omega \partial_b \omega G^{ab}
  +V(\omega)
  +\mathcal{L}_{hd}
  \right]
  \nonumber\\&\qquad\qquad
  +
  \frac12
  M_5^3\int d^4 x dz \,
  \sqrt{g}
  \delta (z-z_{IR})
  \left[
  \tau-V_{brane}(\omega)
  +\mathcal{L}_{brane,hd}
  \right]
  \,,
\end{align}
where $\mathcal{L}_{hd}$ are higher-dimensional operators, suppressed
by the scale $\sim 1/\bar{\rho} $. The pNGB nature of $\omega$ ensures
that $\omega \to \omega+a$ symmetry is only broken by terms suppressed
by $\epsilon$.
Therefore, for $\omega=constant$, the contributions to the bulk action
$V(\omega) \sim \mathcal{O}(\epsilon)$, as well as $\mathcal{L}_{hd}(\omega) \sim
\mathcal{O}(\epsilon)$.
There is no such restriction on the brane terms (except the
$\eta$-tuning on the brane that ensures we can work in perturbation
theory with small backreaction).

Note that this form of the 5D effective action relies only on the
symmetries in the 5D theory, and therefore is valid for a general 6D
action respecting those symmetries. In particular, the simplifying
assumption of a circularly symmetric solution in 6D is not necessary,
and adding deformations away from the circular symmetry do not affect the low
energy effective theory.
This is expected from the fact that there
are no light degrees of freedom associated with this circular
symmetry (appendix \ref{sec:zeromodes}).

We have already derived the effective potential for the dilaton at
leading order in $\epsilon$ for this Lagrangian in
section~\ref{sec:sol5d}. The only difference in
this case is the presence of generic potential and higher dimensional
operators. 
However, these merely correct the form of the leading 
potential, without affecting the $\epsilon$-suppression of the GW
field potential, and hence the $\beta$-function. Since our 4D solution
was general, we conclude that these subleading effects cannot affect
the conclusion.

Higher derivative operators need to be treated with some care in the
presence of branes \cite{Lewandowski:2001qp}. Using equations of motion,
bulk higher derivative operators can be cast in a form such that any
$\omega$ only has at most one derivative acting on it. Such operators
yield well-defined expressions in perturbation theory in $\eta$. The
brane terms need classical renormalization, but the stabilization of
the 5D radius does not depend on the details of the renormalization.

The radius stabilization calculation then proceeds as before. In
particular, as we can see from the 6D solution that one can work
perturbatively in $\eta$. At $\epsilon=0$, we recover pure AdS
solution away from the IR brane, indicating that the CFT is not
explicitly broken
and the radion potential needs to be fine tuned. This is a consequence
of the fact the $\omega=const.$ is a solution to the e.o.m in the
bulk. Then, a maximally symmetric solution of 5D exists, i.e. the bulk
is AdS. 

If we turn on $\epsilon$,  the solution close to the IR brane is still
dominated by the $\epsilon=0$ solution. The $\beta$ function depends
on the potential, which is always suppressed by $\epsilon$. 
Away from the brane, $\omega$ evolves
slowly since its potential is small. Consequently, higher derivative
operators are suppressed with even higher powers of 
the $\epsilon$. Therefore, the solution for $\omega$ found explicitly
in the 6D case holds more generally. 

As noted in \cite{Lewandowski:2001qp, Coradeschi:2013gda,
Chacko:2013dra}, in the presence of
higher dimensional operators, the perturbation theory in $\eta$ we
used to derive the solution is not valid for
$\eta > \epsilon$. Crucially, one needs to invoke a shift symmetry for
the GW field in order to ensure that higher order corrections in
$\eta$ are also higher order in $\epsilon$. The general form of the
perturbative expansion in 5D can be found in
\cite{Lewandowski:2001qp}, where it was shown in detail that in the
presence of a shift symmetry for the GW field, higher order
corrections in both $\epsilon$ and $\eta$ are subdominant. Therefore,
the dilaton potential derived in section \ref{sec:sol5d} is the
leading contribution.

\section{Conclusion and discussion}
\label{sec:conc}
It is very surprising to find a non-supersymmetric CFT in a
spontaneously broken phase, and is usually associated with tuning.
This tuning can be identified with a tuning of the cosmological
constant in a scalar theory of gravity. While spontaneous breaking is
non-generic, if there exists
a deformation with specific properties, then it is possible to obtain
a
low-energy theory that has 
an approximate scale symmetry.
With such an appropriately chosen deformation, the dilaton has an
approximate shift symmetry around its minimum. The condition required
to achieve a light
dilaton are stated most simply in the 5D formulation of the theory. In
order
to get a light dilaton, one requires
an approximate shift symmetry for the stabilizing Goldberger-Wise
field, making its potential
extremely small. 
In this paper we provide a partial UV completion to obtain the
shift symmetry from a higher dimensional gauge field, which
consequently can be completely consistent with quantum gravity
expectations. It will be interesting to see if our solution can be
embedded in to a full string theory UV completion. This is beyond the
scope of the current paper.

It would be interesting to translate our full $A_6$ mechanism into 4D
QFT language, and identify what is the dual to the mechanism that
protects the dilaton potential. That could lead to explicit
constructions in 4D language which would be very interesting for study
of non-supersymmetric CFTs.

In this paper we have studied the vacuum structure of the scalar
gravity cosmological
constant problem. A full solution to the problem must also address
cosmological evolution including obtaining a hot big bang. We will
study this in future work. The interplay of cosmological evolution and
naturalness also appears in the relaxion proposal for solving the
hierarchy problem~\cite{Graham:2015cka}, although it takes a somewhat
different form.

Needless to say, it would be very interesting if some sort of modified
spin-2 gravity with very small violations of the equivalence principle
could yield a naturally small cosmological constant in a manner
analogous to the spin-0 mechanism (references~\cite{Coradeschi:2013gda,Bellazzini:2013fga} explored ideas
in this direction).
 This will require something more
dramatic since Lorentz symmetry guarantees the equivalence principle
for a massless spin-2 particle.

\acknowledgments
We would like to thank Zackaria Chacko, Riccardo
Rattazzi and Matthew Strassler for helpful discussions. The research
of PA is supported by NSF grant PHY-1216270. 
 R.S.~is supported in part by the NSF under Grant
No. PHY-1315155 and by the Maryland Center for Fundamental Physics. 
PA would like to
thank the Aspen Center for Physics where part of this work was
completed.  
\appendix

\section{Aharanov-Bohm potential}
\label{app:casimir}
The Casimir energy contribution from a charged boson for a
$d$-dimensional theory with 1 dimension compactified on a circle of
radius $\rho$ is, (following e.g.~\cite{ArkaniHamed:2007gg}),
\begin{align}
  V(A_6)
  &=
  -2\pi \rho
  \sum_{n=1}^{\infty}
  \frac{2m^d}{(2\pi)^{d/2}}
  \frac{\cos(n\theta)}{(2\pi \rho m n)^{d/2}}
  K_{d/2}(2\pi \rho m n)
  \,,
\end{align}
where $\theta = \oint_{S_1} e A_6=2\pi e A_6$ and $m$ is the mass of
the corresponding charged field. The fermionic contribution has an
additional overall negative sign, and anti-periodic boundary
conditions yield a $n$-dependent negative sign.

For the bulk potential in our case, this becomes
\begin{align}
  V_{bulk}(A_6)
  &=
  -\frac{2\pi \rho}{M_6^4}
  \sum_{n=1}^{\infty}
  \frac{2m_\xi^6}{(2\pi)^3}
  \frac{\cos(2\pi n e A_6)}{(2\pi \rho m_\xi n)^3}
  K_3(2\pi \rho m_\xi n)
  \,,
\end{align}
where we have accounted for our normalization of the action by
$M_6^4$. 
Since we are interested in the limit $2 \pi \rho m_\xi \gg 1$, we can
use the
asymptotic behavior of the Bessel function,
\begin{align}
  K_\nu(x)
  \xrightarrow{x\to\infty}
  \sqrt{\frac{\pi}{2 x}}
  e^{-x}
  \,.
\end{align}
Therefore, the dominant contribution comes from $n=1$,
\begin{align}
  V_{bulk}(A_6)
  &\simeq
  \frac{1}{M_6^4}
\left(\frac{m_\xi}{4\pi^2 \rho} \right)^{5/2}
  e^{-2\pi \rho m_\xi }
  \cos(2\pi \,e A_6)
  \,.
\end{align}

Since this is an exponentially small number, one worry is if
non-perturbative effects can overcome this suppression.
This issue was considered in e.g.~\cite{ArkaniHamed:2007gg}, with the
conclusion that non-perturbative effects are also similarly
exponentially suppressed.

For the brane potential, we can approximate the potential in the
massless limit $m \rho \ll 1$,
\begin{align}
  V_{brane}
  &=
  -\frac{1}{3\sqrt{2}(2\pi)^2\rho^4 M_6^4}
    \sum_k \frac{e^{2\pi i k e A_6}}{k^5}
    +h.c.
\end{align}
Ignoring the higher harmonics which are suppressed,
\begin{align}
  V_{brane}
  &=
  -\frac{\sqrt{2}}{3(2\pi)^2\rho^4 M_6^4}
    \cos(2\pi e A_6)
    \,.
\end{align}
Multiple charged fields will give rise to a sum of such terms,
resulting in a generic brane potential.

\section{Relevant scales}
In this section we outline the (mild) hierarchy of scales that we have
assumed.
In order to have a well-defined 6D effective field theory, we want
the 6D Planck scale $M_6$ to be somewhat larger than the
inverse size of the extra-dimension. Further, since we want a
moderately heavy charged particle in the 6D theory $m_\xi$, this
should be captured within the EFT as well,
\begin{align}
  M_6 > m_\xi > \frac{1}{\bar{\rho}} 
  \,.
\end{align}
Similarly, for the 5D effective field theory, we require,
\begin{align}
   M_5 > k_5
  \,.
\end{align}
Thus, the hierarchy of scales is,
\begin{align}
   M_5 > M_6 > m_\xi > \frac{1}{\bar{\rho}}> k_5 \gg M_{pl}
\end{align}
where $M_{pl}$ is the 4D scalar gravity Planck scale.
We can write these inequalities in terms of Lagrangian parameters in
the 6D theory, $\{k_5,v, m_\xi\}$ using
\begin{align}
  \frac{1}{\bar{\rho}} 
  &=
  \frac{k_5}{v}
  \\
  M_5 
  &= 
  (2\pi \bar{\rho} M_6^4)^\frac13
  M_6
  =
  \left(
  \frac{2\pi v M_6}{k_5}
  \right)^\frac13
  M_6
  \,.
\end{align}
We can check that these inequalities are satisfied for
\begin{align}
  m_\xi< M_6\,,
  \quad
  v<1\,,
  \quad
  k_5 < \frac{M_6}{\mathcal{N}} 
  \,,
\end{align}
where $\mathcal{N} = \bar{\rho} m_\xi \sim100$ in order to get 
$\epsilon \sim 10^{-120}$. We see that the 6D theory has only mild
hierarchies.

\section{Circular symmetry in 6D}
\label{sec:zeromodes}
In this section we show that the $U(1)$ symmetry used for simplifying
calculations in section~\ref{sec:sol6d} does not have any light
degrees of freedom associated with it and hence does not appear in the
5D effective theory. Consequently, departure from the circularly
symmetric
solution in 6D will not modify the general form of the 5D effective
theory.

There are potentially two gravitational
massless excitations: the Kaluza-Klein $U(1)$ gauge field
associated with the $S_1$ compactification, and the excitation
$\sigma(x,z,\vartheta)=\vartheta+\tilde\sigma(x,z)$.
Clearly, any configuration $\tilde\sigma(x,z)$ can be absorbed by a
$(x,z)$-dependent co-ordinate rotation in the sixth dimension
$\vartheta$. In other words, $\tilde\sigma(x,z)=0$ defines the unitary gauge
condition for the KK U(1) gauge boson. Thus, KK U(1) is spontaneously
broken, with the gauge boson acquiring a mass. The symmetry breaking
pattern is $U(1)_{KK}\times U(1)_{global,\chi} \to
U(1)_{global,\chi}$, and hence no massless Goldstone appears.

This can be seen by substituting the following ansatz back into the
action, 
\begin{align}
ds^2
&=
e^{-2k_5 z} dx^\mu dx^\nu \eta_{\mu\nu}
-dz^2 - (\rho d\vartheta - \sqrt{2}V_M dx^M)^2
\,,
\quad 
\chi = v e^{i\vartheta} 
\,,
\end{align}
yielding
\begin{align}
  S
  &=
  M_6^4\int d^4 x \, dz 
  \sqrt{g}\, (2\pi \rho)
  \left[
    -\frac14 R[g] +4k_5^2 -\frac14 V_{MN} V_{AB} g^{MA}g^{NB} 
    + \frac{v^2}{\rho^2} 
    \left(-1 +2 V_A V_B g^{AB} \right)
  \right]
  \,.
\end{align}
We see that the KK $U(1)$ gauge boson does obtain a mass of order the 5D
curvature scale.
\\

\bibliography{scalar-gravity.bib}

\bibliographystyle{JHEP.bst}
\end{document}